\documentclass[reprint,aps,superscriptaddress,amsmath,amssymb,longbibliography]{revtex4-1}
                                                                                                                                                                                                                                                                                                                                                                                                                                                                                                                                                                                                                                                                                                                                                                                                                                                                                                                                                                                                                                                                                                                                                                                                                                                                                                                                                                                                                                                                                                                                                                                                                                                                                                                                                                                                                                                                                                                                                                                                                                                                                                                                                                                                                                                                                                                                                                                                                                                                                                                                                                                                                                                                                                                                                                                                                                                                                                                                                                                                                                                                                                                                                                                                                                                                                                                                                                       
\usepackage{graphicx}
\usepackage{dcolumn}
\usepackage{bm}
\usepackage{color}
\usepackage{xspace}
\usepackage{ulem}
\usepackage{xfrac}
\usepackage{soul}
\usepackage[colorlinks=true,urlcolor=blue,citecolor=blue,linkcolor=blue,bookmarks=false,
pdfstartview={FitH}]{hyperref}

\def \cm-1{cm$^{-1}$\,}
\def\FTS{Fe$_{1+y}$Te$_{0.6}$Se$_{0.4}$\,} 
\def\FTSxy{Fe$_{1+y}$Te$_{1-x}$Se$_{x}$\,}
\def\cm{cm$^{-1}$}

\def\Tc{$T_c$}
                                                                                                                                                                                                                                                                                                                                                                                                                                                                                                                                             
\begin{document} 
                                                                                                                                                         
\title{Superconductivity and phonon self-energy effects in \FTS}

\author{S.-F. Wu}\email{sw666@physics.rutgers.edu}
\affiliation{Department of Physics and Astronomy, Rutgers University,
Piscataway, NJ 08854, USA}
\author{A. Almoalem}
\affiliation{Physics Department, Technion-Israel Institute of Technology, Haifa 32000, Israel}  
\author{I. Feldman}
\affiliation{Physics Department, Technion-Israel Institute of Technology, Haifa 32000, Israel}  
\author{A. Lee}
\affiliation{Department of Physics and Astronomy, Rutgers University,
Piscataway, NJ 08854, USA}    
\author{A. Kanigel}
\affiliation{Physics Department, Technion-Israel Institute of Technology, Haifa 32000, Israel}   
\author{G. Blumberg}\email{girsh@physics.rutgers.edu}
\affiliation{Department of Physics and Astronomy, Rutgers University,
Piscataway, NJ 08854, USA}
\affiliation{National Institute of Chemical Physics and Biophysics,
12618 Tallinn, Estonia}
\date{\today}                 
                                                                                                                                                                                                                                                                                                                                                                                                                                                                                                                                                                                                                                                                                                                                                                                                                                                                                                                                                                                                                                                                                                                                                                                                                                                                                                                                                                                                                                                         
\begin{abstract}                                          
We study \FTS multi-band superconductor with $T_c=14$\,K by polarization-resolved Raman spectroscopy.  
Deep in the superconducting state, we detect pair-breaking excitation at 45\,\cm($2\Delta=5.6$\,meV) in the $XY$($B_{2g}$) scattering geometry, consistent with twice of the superconducting gap energy (3\,meV) revealed by ARPES on the hole-like Fermi pocket with $d_{xz}/d_{yz}$ character.  
We analyze the superconductivity induced phonon self-energy effects for the $B_{1g}$(Fe) phonon and estimate the electron-phonon coupling constant $\lambda^\Gamma \approx 0.026$, which is insufficient to explain superconductivity with  $T_c=14$\,K.                                                                                                                                                                                                   
\end{abstract}
                                                                                                                                                                                                                                                                                                                                                                                                                                                                                                                                                                                                                                                                                                                                                                                                                                                         
                                                                                                                                                                                                                                                                                                                                                                                                                                                                                                                                                                                                                                                                                                                                                                                                
\maketitle                                                                                                                                                                                                                                                                                         
\textit{Introduction}
-- Since the discovery of the multi-band iron-based superconductors (FeSCs) in 2008~\cite{Kamihara2008JACS}, a unified understanding the pairing mechanism in FeSCs remains in a focus of attention~\cite{Paglione2010NatPhy,Wang2011science,Fernandes_Chubukov_review,Hirschfeld2016review,Dai2015RevModPhys,Si2016NSR,Coleman2019arxiv}. 
One step towards understanding the pairing mechanism is to study the superconducting (SC) gaps on different pockets of the Fermi surface (FS)~\cite{Richard_2015,Hirschfeld2016review}.

\begin{figure}[!t] 
\begin{center}
\includegraphics[width=\columnwidth]{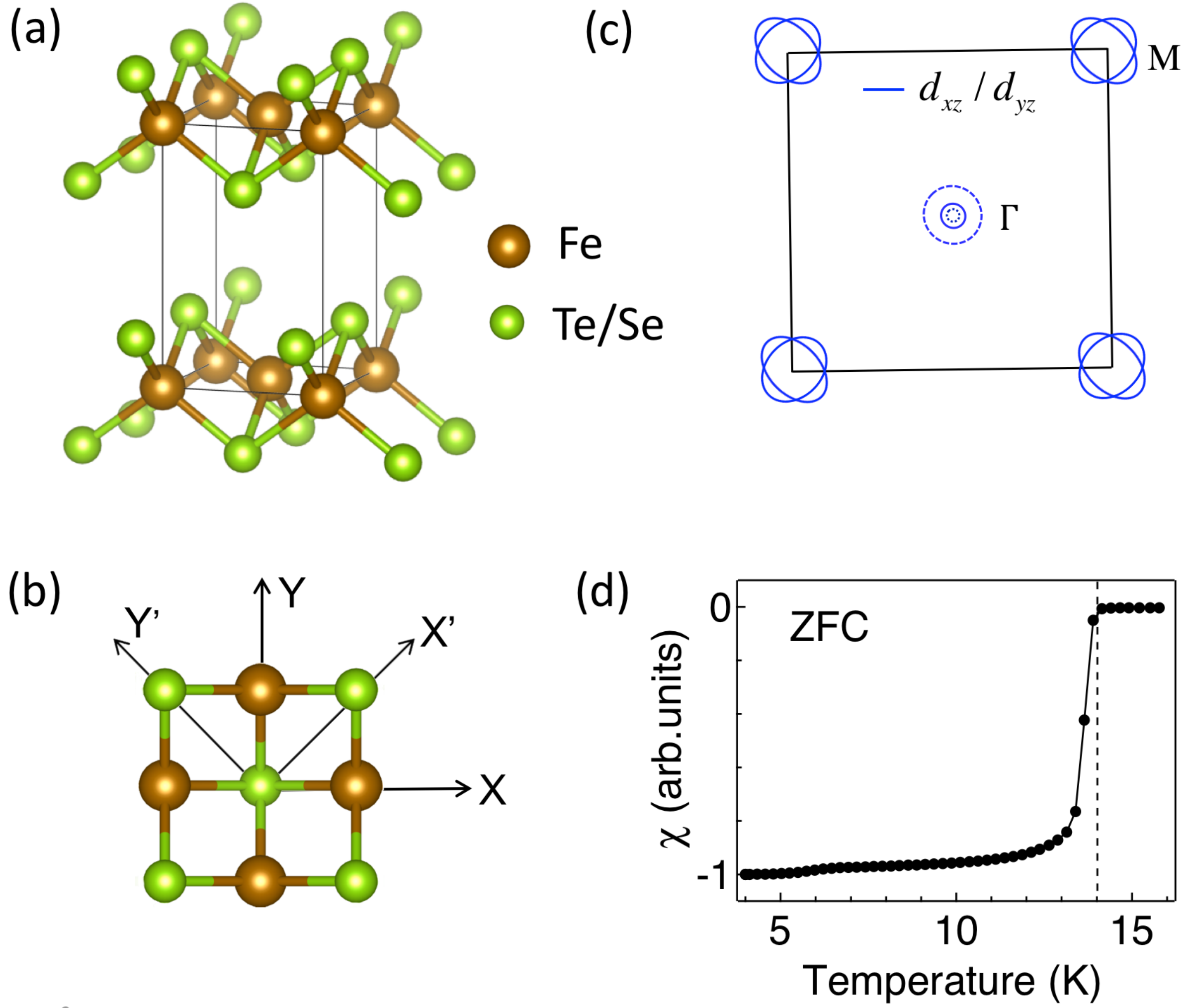}
\end{center}
\caption{\label{Fig1_intro} 
(a) Crystal structure of \FTS. 
(b) Definition of $X$, $Y$, $X'$ and $Y'$ directions in a 2-Fe unit cell. 
(c) Schematic representation of Fermi surfaces of \FTS in the 2-Fe Brillouin zone (BZ)~\cite{dxynote}.
(d) "Zero-field-cool" magnetic susceptibility measured with magnetic field H=4\,Oe along the $c$-axis direction.}
\end{figure}

The chalcogenide family Fe$_{1+y}$Te$_{1-x}$Se$_x$ has simple stoichiometry and crystal structure which can be viewed as stacks of FeTe$_{1-x}$Se$_x$ layers~[Fig.~\ref{Fig1_intro}(a)]. 
Superconductivity in this system was first found at 8\,K in the nonmagnetic FeSe. 
With about 60\% isovalent substitution of Se for Te, 
$T_c$ in \FTS increases to 14\,K~\cite{Fang2008PhysRevB}. 
Thus, the non-magnetic and tetragonal \FTS is an ideal system to study the SC order parameter, without the effect of coexisting or interacting with other electronic orders.
                    
Polarization-resolved Raman spectroscopy has been used to study the  pair-breaking excitations in different symmetry channels~\cite{Muschler2009PRB,chauviere2010PRB,Massat2016PNAS,Sugai2010PhysRevB,Rudi_2014PRX,Wu2017PhysRevB,Thorsmolle2016PRB} and for estimation of the electron-phonon coupling~\cite{Choi_2010_JPCM,Um2014PRBNFCA,Zhang_2018PhysRevB} in FeSCs.    
Previous Raman studies on \FTSxy were focused on the lattice dynamics~\cite{Xia_2009PRB,Gnezdilov_PRB83,Um_2012PRB,Okazaki_2011PRB} and the magnetic excitations~\cite{Okazaki_2011PRB}, while the superconducting features for \FTS have not been well established.              
                                                                                                                                                                                                                                                                                                                                                                                                                                                                                                                                                                                                                                                                                                                                                                                                                                                                                                                                                                                                                   
In this work, we use polarization-resolved Raman spectroscopy to study the pair breaking excitations in \FTS. 
In the SC state, we identify the coherence peak at 45\,\cm($2\Delta=5.6$\,meV) in the $XY$($B_{2g}$ in $D_{4h}$) scattering geometry with cross-polarized light along Fe-Te/Se directions. 
The peak energy is close to the twice of the gap energy (3\,meV) on the hole-like FS pocket with $d_{xz}/d_{yz}$ character around $\Gamma$ point, as determined by ARPES.
We investigate the superconductivity induced phonon self-energy effects for the $B_{1g}$(Fe) phonon mode and estimate the electron-phonon coupling constant $\lambda^\Gamma \approx 0.026$, which is very weak to explain $T_c=14$\,K in \FTS.

\textit{Experimental}\label{Experiment and Methods}                                            
-- The single crystals of \FTS were grown using modified Bridgman method~\cite{Rinotte2017SciAdv}. 
The composition of \FTS was determined on samples from the same growth batch by energy-dispersive x-ray (EDX) analysis. 
The nominal composition of excess Fe ($y$) in \FTS is about 2\%, while the real value of $y$ is close to zero~\cite{Rinotte2017SciAdv}. 
The magnetic susceptibility confirming a sharp SC transition at $T_c$=14\,K is shown in Fig.~\ref{Fig1_intro}(d). 
                                                                                                                                                                   
The \FTS single crystals used for Raman measurements were cleaved in nitrogen gas atmosphere in a glove bag, then immediately loaded into connected continuous helium flow optical cryostat and quickly cooled down to 250\,K within 5 minutes to avoid surface contamination. 
All the Raman scattering measurements were performed using the Kr$^+$ laser line at 647.1\,nm (1.92\,eV) in a quasi-back scattering geometry along the crystallographic $c$-axis with an instrumental resolution about 1.5\,\cm.
The excitation laser beam was focused into a $50\times100$ $\mu$m$^2$ spot on the $ab$-surface, with the incident power around 10\,mW and 2\,mW for normal state and superconducting state measurements, respectively. 
The scattered light was collected and analyzed by a triple-stage aberration corrected Raman spectrometer and recorded using a liquid nitrogen-cooled charge-coupled detector. 
The Raman spectra were corrected for the spectra response of the spectrometer and the detector. 
The temperature shown in this paper has been corrected for laser heating. 
An empirical heating coefficient 1\,K/mW was applied according to FeSe~\cite{Massat2016PNAS,Kretzschmar2016NatPhy,Weilu_FeSe2017arXiv}. 
                                                                                                                                                                       
In this manuscript, we define the $X$ and $Y$ directions along the two-Fe unit cell basis vectors (at 45$^{\circ}$ degrees from the Fe-Fe directions) in the tetragonal phase, whereas $X'$ and $Y'$ are
along the Fe-Fe directions~[Fig.~\ref{Fig1_intro}(b)].
Raman spectra were recorded for ($\hat{e}_i \hat{e}_s$) = ($XX$), ($XY$), ($X'X'$) and ($X'Y'$) polarization geometries in the $ab$-plane, where $\hat{e}_i$ and  $\hat{e}_s$ represent the incident and scattered light polarization, respectively.   
                                                                                                                                                                                                                                                                                                        
For crystals with point group symmetry $D_{4h}$, the $XX$, $X^{\prime}Y^{\prime}$ and $XY$ geometries probe $A_{1g}$+$B_{1g}$, $A_{2g}$+$B_{1g}$ and $A_{2g}$+$B_{2g}$ channels, respectively~\cite{Devereaux2007RMP}. 
Assuming the $A_{2g}$ response is negligible~\cite{Muschler2009PRB} and using the background estimated from the $X'Y'$-symmetry electronic continuum~\cite{Thorsmolle2016PRB,Gallais2013PRL}, the Raman response in $(\mu\,\nu)$ scattering geometry  $\chi''_{\mu \nu}(\omega,T)$ can be obtained: $\chi''_{\mu \nu}(\omega,T)=I_{\mu \nu}(\omega,T)/[1+n(\omega,T)]$, where $I_{\mu \nu}(\omega,T)$ is the Raman intensity after background subtraction and $n(\omega,T)$ is the Bose-Einstein factor. 
                                                                                                                                                                                                                                                                                                                                              
\textit{The normal state}\label{Normal_state} -- 
The \FTS crystal structure belongs to the space group $P4/nmm$ (point group $D_{4h}$). 
The $\Gamma$ point Raman active modes are $\Gamma_{Raman}$ = $A_{1g}$ + $B_{1g}$ + 2$E_{g}$. 
The $A_{1g}$ and $B_{1g}$ modes are related with Te/Se and Fe atoms $c$-axis lattice vibrations, respectively. 
                                                                                                                                                                                                                                                                                                                                                                                                                                                                                                                                                                                                                                                                                                                                           
\begin{table}[!b]
\caption{\label{phonon} 
Comparison of the phonon frequencies ($\omega$) and the line-widths ($\gamma$) for $A_{1g}$(Te/Se) and $B_{1g}$(Fe) modes for \FTSxy with similar compositions at 5\,K. Units are in \cm.}
\begin{ruledtabular}
\begin{tabular}{ccccc}
Sample&$\omega_{A_{1g}}$&$\gamma_{A_{1g}}$&$\omega_{B_{1g}}$&$\gamma_{B_{1g}}$\\
\hline
FeTe$_{0.6}$Se$_{0.4}$~\cite{Okazaki_2011PRB} &162.6&27.2~\cite{Linewidth}&207.7&8.8~\cite{Linewidth}\\
\hline
Fe$_{0.95}$Te$_{0.56}$Se$_{0.44}$~\cite{Um_2012PRB} &162&20&207&5\\
\hline
\FTS(this work)&162&6.9&208.2&4\\
\end{tabular}
\end{ruledtabular}
\begin{raggedright}
\end{raggedright}
\end{table}

\begin{figure}[!t] 
\begin{center}
\includegraphics[width=\columnwidth]{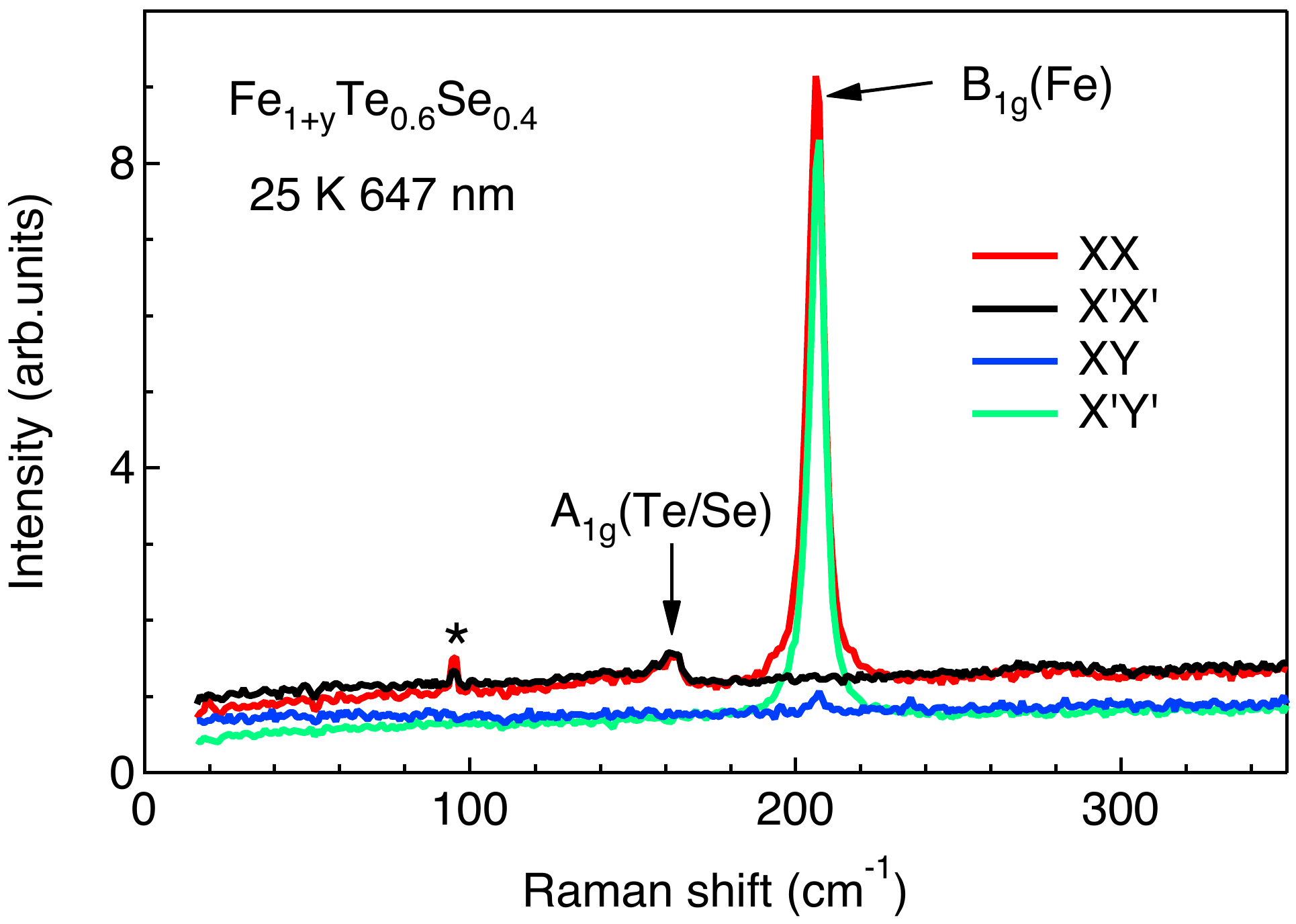}
\end{center}
\caption{\label{Fig2_15K_4pol} Raman intensity of \FTS at 25\,K for $XX$, $XY$, $X'X'$ and $X'Y'$ scattering geometries. The star represents a laser plasma line.
}
\end{figure}

In Fig.~\ref{Fig2_15K_4pol}, we show the normal state Raman spectra for four in-plane scattering geometries at 25\,K. 
The observed mode at around 162\,\cm\ and the mode at around 207\,\cm\ in $XX$ scattering geometry are assigned to $A_{1g}$(Te/Se) and $B_{1g}$(Fe) modes, respectively~\cite{Okazaki_2011PRB,Um_2012PRB}. 
The comparison of the phonon frequencies and line-widths obtained in this work with previous studies~\cite{Okazaki_2011PRB,Um_2012PRB} are summarized in Table~\ref{phonon}. 
The phonon frequency for both modes and the line-width for the $B_{1g}$(Fe) mode are consistent with Ref.~\cite{Okazaki_2011PRB,Um_2012PRB}. 
In contrast, the line-width for the $A_{1g}$(Te/Se) mode in this work is about 3$\sim$4 times sharper than in Ref.~\cite{Okazaki_2011PRB,Um_2012PRB}.
We note that the spectra shown in Fig.~\ref{Fig2_15K_4pol} lack the two features reported in Ref.~\cite{Okazaki_2011PRB}: the low frequency bump/tail appearing in all in-plane scattering geometries and a substantial `leakage' intensity of $A_{1g}$ phonon into cross scattering geometries. 
The absence of these two features indicates the quality of cleaved surface in this study.  
                                
%
                                                                                                                                                                                                                                                                                                                                                                                                                                                                                                   
\begin{figure}[!t] 
\begin{center}
\includegraphics[width=\columnwidth]{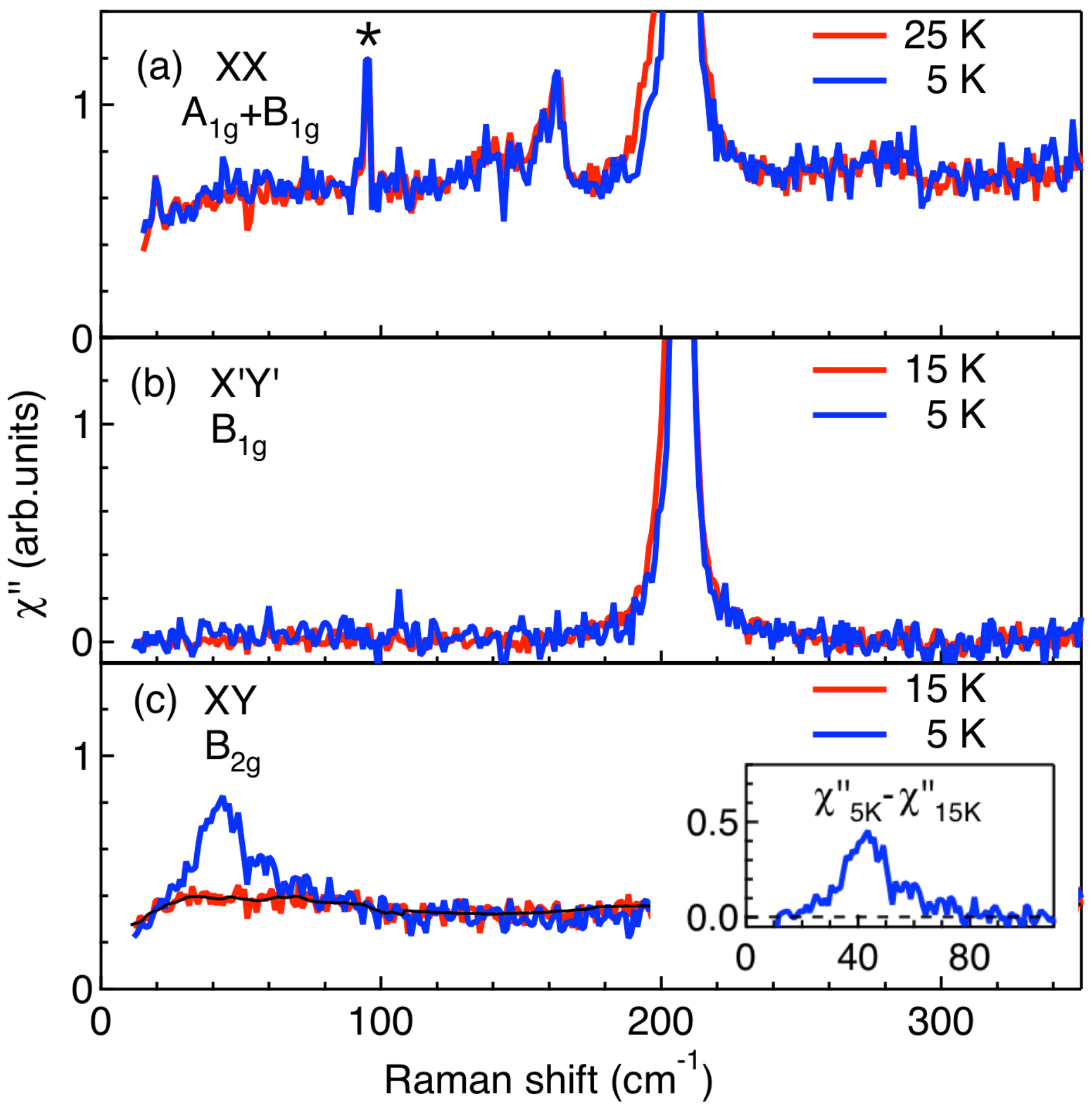}
\end{center}
\caption{\label{Fig3_SC} Raman response of \FTS above and below \Tc\ in $XX$ (a), $X'Y'$ (b) and $XY$ (c) scattering geometries. 
The star in (a) indicates a laser plasma line.
The solid black line in (c) represent the smoothed curve for the Raman response at 15\,K which is used to calculate the difference between the response above and below \Tc\ in $XY$ scattering geometry shown in the inset of (c).}
\end{figure}     

\textit{The superconducting state}\label{SC} 
-- Before looking into the Raman features observed in the SC state for \FTS, we recall the SC gap values obtained by complementary spectroscopic probes.
Angle-resolved photoemission spectroscopy (ARPES) revealed nodeless SC gaps with small or negligible in-plane anisotropy on both hole-like and electron-like FS pockets\,\cite{Miao2012PhysRevB,Rinotte2017SciAdv,Zhang_science2018}. 
In addition to the SC gap of 1.8\,meV for the topological surface state around the $\Gamma$ point~\cite{Zhang_science2018}, two close to isotropic gap values for bulk bands with $d_{xz}/d_{yz}$ character were reported: 
a 3\,meV gap for the hole-like FS pocket centered around the $\Gamma$ point~\cite{Rinotte2017SciAdv} and a larger, about 4.2\,meV gap for the electron pocket around the M point~\cite{Miao2012PhysRevB}.         
The scanning tunneling spectroscopy (STS) showed SC coherence peaks at 1.5\,meV and a shoulder at 2.5\,meV~\cite{Yin2015NatPhy}. 
The infrared spectroscopy (IR) revealed two SC gaps at 2.47\,meV and 5.08\,meV~\cite{Homes2010PhysRevB}. 
The measurements of specific heat  confirmed the nodeless nature of the SC order parameter with an averaged gap value 2.3\,meV~\cite{Escudero2015}. 
At the energy scale of the sum of SC gaps on the hole-like and electron pockets with $d_{xz}/d_{yz}$ character, a 6.5\,meV neutron-spin-resonance mode was reported below $T_c$ at $\mathbf{Q}=(\pi,\pi)$ [2-Fe BZ] by inelastic neutron scattering (INS) measurements~\cite{Qiu2009PhysRevLett}. 
The values of the SC gaps and bosonic modes deduced from different spectroscopies are summarized in Table~\ref{gap}.    
                                                                                                                                                                                                                                                                                                                                           
\begin{table}[!b]
\caption{\label{gap} The summary of the values for SC gaps and bosonic modes deducted from ARPES, STS, IR, Specific heat, Raman scattering, and INS measurements for \FTS with similar compositions. 
$\Delta_{1}$, $\Delta_{2}$ and $\Delta_{3}$ represent three energy scales of gap values. 
$E_{CM}$ represents the energy of the bosonic mode. 
All energies are given in units of meV.}
\begin{ruledtabular}
\begin{tabular}{cccccc}
&$\Delta_{1}$&$\Delta_{2}$&$\Delta_{3}$&$E_{CM}$\\
\hline
ARPES&1.8~\cite{Zhang_science2018}&3~\cite{Rinotte2017SciAdv}&4.2~\cite{Miao2012PhysRevB}&\\
STS&1.5~\cite{Yin2015NatPhy,Hanaguri2010science}&2.5~\cite{Yin2015NatPhy}&&\\
IR&&2.47~\cite{Homes2010PhysRevB}&5.08~\cite{Homes2010PhysRevB}&&\\
Specific heat&&2.3~\cite{Escudero2015}&\\
Raman&&2.8 (this work)& &\\
INS&&& &6.5~\cite{Qiu2009PhysRevLett}&\\
\end{tabular}
\end{ruledtabular}
\begin{raggedright}
\end{raggedright}
\end{table}                                                                                                                                                                                      

In Fig.~\ref{Fig3_SC}, we present the Raman response above and below \Tc\ for three scattering geometries. 
For the $XX$ and $X^\prime$Y$^\prime$ scattering geometries, the electronic continuum barely changes upon cooling below \Tc. 
In contrast, for $XY$ scattering geometry a clear peak at around 45\,\cm\,, which we relate to the pair-breaking excitation,  emerges below \Tc. 
Since the peak's energy is about 5.6\,meV, based on the Table~\ref{gap}, we assign it to the gap 2$\Delta_{2}$ on the hole-like FS pocket with the $d_{xz}/d_{yz}$ character~\cite{Zhang2019NatPhy} around $\Gamma$ point with a typical gap value $\Delta_{2}=3$\,meV from the ARPES measurement~\cite{Rinotte2017SciAdv}. 
The pair-breaking peak associated with the larger gap 2$\Delta_{3}=8.4$\,meV on the electron pockets around $M$ point is not detected in the $B_{2g}$ channel. 
In contrast, in optimally-doped Ba(Fe$_{0.939}$Co$_{0.061}$)$_2$As$_2$, a similar single peak observed in $B_{2g}$ channel was interpreted as a pair-breaking peak originated from the electron pockets around $M$ point~\cite{Muschler2009PRB}. 
While \FTS and Ba(Fe$_{0.939}$Co$_{0.061}$)$_2$As$_2$ share similar FS topology, these differences suggest that the Raman vertex in the $B_{2g}$ channel for multi-band FeSCs is rather complex.
                                                                                                                                                                                                                          
\begin{figure}[!t] 
\begin{center}
\includegraphics[width=\columnwidth]{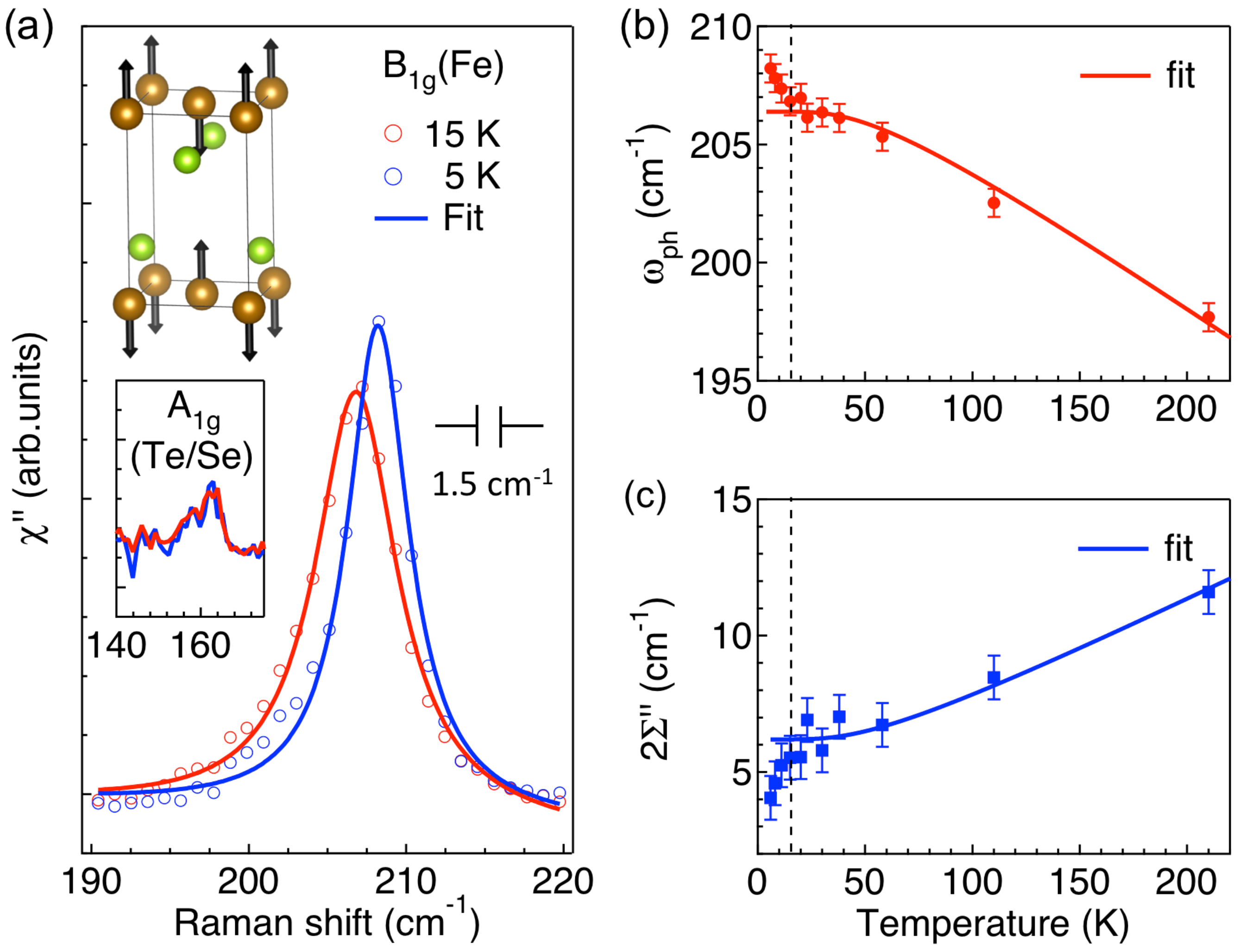}
\end{center}
\caption{\label{Fig4_B1g_phonon} 
(a) $B_{1g}$(Fe) phonon for \FTS at 15\,K and 5\,K in $X'Y'$ scattering geometry. The top inset of (a) illustrates the atomic displacement of $B_{1g}$(Fe) phonon. 
The bottom inset of (a) shows the $A_{1g}$(Te/Se) phonon at 25\,K and 5\,K in $XX$ scattering geometry. 
(b)-(c) T-dependence of the $B_{1g}$(Fe) phonon energy $\omega_{ph}$ and line-width $2\Sigma''$. 
$\Sigma''$ has been corrected for the instrumental resolution.
The solid lines in (b) and (c) represent the fitting of the normal state phonon by anharmonic decay model~\cite{B1g_phonon_note}. 
The dashed vertical lines in (b) and (c) represent $T_c$.}
\end{figure}

\textit{Phonon self-energy effects}
-- The frequency and line-width for the $B_{1g}$(Fe) phonon are presented in Fig.~\ref{Fig4_B1g_phonon}(b-c). 
Above $T_c$, the $B_{1g}$(Fe) phonon shows a conventional temperature dependence: hardening and sharpening upon cooling following the anharmonic phonon decay  model~\cite{Klemens_PhysRev148,Menendez_PRB29,B1g_phonon_note}. 
Below $T_c$, in contrast to the $A_{1g}$(Te/Se) phonon which barely changes upon cooling across $T_c$~[inset of Fig.~\ref{Fig4_B1g_phonon}(a)], the $B_{1g}$(Fe) mode shows abnormal behavior in the SC state. The mode's energy and line-width show additional hardening and sharpening in the SC state [Fig.~\ref{Fig4_B1g_phonon}(a)].
                                                                                                                                                                                                                                                                                                                                                                                                                                                                                      
The $B_{1g}$(Fe) phonon energy is around 26\,meV. It is much larger than the twice of the maximum gap $2\Delta_{3}$=8.4\,meV in \FTS~\cite{Miao2012PhysRevB}. 
For a phonon with $\omega > 2 \Delta$, such hardening upon entering into SC state was also reported for MgB$_2$~\cite{Mialitsin_PRB2007}, for cuprate
superconductors~\cite{Cooper_PRB1988,Thomsen_PRB1988,Friedl_PRL1990,McCarty_PRB1991,blumberg1994JSC,Thomsen_1998}, and for FeSCs~\cite{Choi_2010_JPCM,Um2014PRBNFCA,Zhang_2018PhysRevB}.
The effects were explained within the Zeyher-Zwicknagl's model~\cite{Zeyher1990} as a consequence of electron-phonon coupling. 
When the SC gap opens below $T_c$, the electronic density-of-states around the Fermi level is reorganized and pushed above the $2\Delta$ energies to the proximity of the phonon energy, then the real part of the phonon self-energy shifts to higher energies due to the renormalization by electron-phonon coupling. 
                                                                                                                                                                                                                                                           
However, the Zeyher-Zwicknagl's model predicts that the line-width for a phonon with $\omega > 2 \Delta$ should be broadened upon entering the SC state as the phonon has additional decay channels. 
To the contrary, the line-width for the $B_{1g}$(Fe) phonon for \FTS further decreases in the SC state. Similar sharpening of the $B_{1g}$(Fe) phonon in the SC state was also observed for NaFe$_{0.97}$Co$_{0.03}$As~\cite{Um2014PRBNFCA}, Ba$_{0.72}$K$_{0.28}$Fe$_2$As$_2$~\cite{Choi_2010_JPCM} and Sr$_{0.85}$K$_{0.15}$Fe$_2$As$_2$~\cite{Choi_2010_JPCM}.
                                                                                                                            
To quantitatively estimate the SC induced self-energy effects and the electron-phonon coupling constant, we use the following model to fit the $B_{1g}$(Fe) phonon:
$\chi_{ph}^{\prime\prime}(\omega)\propto4\omega_0^2\Sigma^{\prime\prime}[(\omega^2-\omega_0^2-2\omega_0\Sigma^\prime)^2+4(\omega_0\Sigma^{\prime\prime})^2]^{-1}$,
where $\omega_0$ is the bare phonon frequency and
$\Sigma=\Sigma^\prime+i\Sigma^{\prime\prime}$ is complex phonon
self-energy~\cite{Zeyher1990}.
The phonon appears at $\omega_{ph}=\sqrt{\omega_0^2+2\omega_0\Sigma^\prime}$ if $\Sigma^{\prime\prime}$ is small. 
The fitting results are presented in Fig.~\ref{Fig4_B1g_phonon}(b-c)~\cite{B1g_phonon_note}.
                                                                                                                                                                                                                                                                                                                                                                                                                                               
We compute the electron-phonon coupling constant $\lambda_{B_{1g}}^\Gamma$ for the $B_{1g}$(Fe) phonon at
$\Gamma$ point~\cite{Rodriguez_PRB1990}: $\lambda=-\kappa\sin u/u$,
where $\kappa=[(\Sigma^\prime(5\,K)-\Sigma^\prime(15\,K))-i(\Sigma^{\prime\prime}(5\,K)-\Sigma^{\prime\prime}(15\,K))]/\omega_{ph}(15\,K)$ and $u\equiv\pi+$
$2i\cosh^{-1}[\omega_{ph}(15\,K)/2\Delta]$. 
With $2\Delta = 45$\,\cm obtained from the pair-breaking peak energy in the $B_{2g}$ channel, we derive  a weak electron-phonon coupling constant
$\lambda_{B_{1g}}^\Gamma \approx 0.026$. 
Since the $A_{1g}$(Te/Se) phonon shows negligible SC-induced self-energy effect [inset of Fig.~\ref{Fig4_B1g_phonon}(a)], $B_{1g}$(Fe) phonon is the only phonon that shows such an effect. Therefore, we can use the $\lambda_{B_{1g}}^\Gamma$ as an approximation for the averaged electron-phonon coupling constant $\lambda^\Gamma$ at BZ zone center~\cite{Zhang_2018PhysRevB}. 

By comparison, a much larger electron-phonon coupling constant $\lambda^\Gamma \approx 0.3$ was reported for a conventional phonon-mediated superconductor MgB$_2$ with $T_c= 39$\,K~\cite{Mialitsin_PRB2007}. 
Furthermore, for V$_3$Ga, an $s$-wave superconductor with a similar $T_c=14.2$\,K, the electron-phonon coupling constant was estimated to be $\lambda \approx 0.9$ by optical measurements~\cite{maksimov1972determination}.
Therefore, the electron-phonon coupling constant $\lambda^\Gamma \approx 0.026$ is clearly insufficient to cause $T_c=14$\,K superconductivity in \FTS.
                                                                                                                            
\textit{Conclusions}
-- In summary, we present polarization-resolved Raman spectroscopic study of 
single crystal \FTS superconductor with $T_c=14$\,K. 
                                                        
In the SC state, we observe a distinct pair-breaking peak at 
45\,\cm($2\Delta=5.6$\,meV) in the $B_{2g}$ channel corresponding to the twice of the gap energy (3\,meV) on the hole-like FS pocket with $d_{xz}/d_{yz}$ character around $\Gamma$ point reported by ARPES measurements.
                                            
We analyze the superconductivity induced phonon self-energy effects for the $B_{1g}$(Fe) phonon and compute the electron-phonon coupling constant $\lambda^\Gamma \approx 0.026$, which is small for \FTS with $T_c=14$\,K.                                
                                                      
\acknowledgments{We thank L.Y.\,Kong and J.X.\,Yin for useful discussions. 
The spectroscopic work at Rutgers (SW, AL and GB) was supported by NSF Grant No. DMR-1709161. 
Crystal growth and characterization at Technion-Israel Institute of Technology (AA, IF and AK) were supported by Israel Science Foundation grant no: 320/17.}
                                                                                                                                                                                                                                                                                                                                                                                                                                                                                                   

\begin{thebibliography}{52}
\bibitem [{B1g()}]{B1g_phonon_note}%
  \BibitemOpen
  \href@noop {} {\ }\bibinfo {note} {Above $T_c$, the temperature dependence of
  the frequency and line-width for the $B_{1g}$(Fe) phonon can be fitted by
  $\omega_{ph}(T)=\omega_{0}-C(1+2/(e^{\hbar\omega_0 /2k_BT} -1)$,
  $\Gamma_{ph}(T)=\Gamma_{0}+\Gamma_1(1+2/e^{\hbar\omega_0/2k_BT} -1)$. The
  fitting parameters for $\omega_{0}$, $C$, $\Gamma_{0}$, $\Gamma_1$ are
  211.1\,\cm, 4.8\,\cm, 3.2\,\cm and 2.9\,\cm, respectively.}\BibitemShut
  {Stop}%
\end{thebibliography}

%

\end{document}